\documentclass[Royal,sageh,times]{sagej}

\usepackage{moreverb,url}

\usepackage{natbib}
\usepackage{graphicx}
\usepackage{enumerate}
\usepackage{url}
\usepackage{amsthm} 
\newtheorem{definition}{Definition}
\usepackage{amsmath} 
\usepackage{xcolor} 
\usepackage{amsfonts} 
\usepackage{tikz}
\def\mm{m}
\usepackage{algorithm}
\usepackage{algpseudocode}

\usepackage{bibunits}

\usepackage[colorlinks,bookmarksopen,bookmarksnumbered,citecolor=red,urlcolor=red]{hyperref}

\newcommand\BibTeX{{\rmfamily B\kern-.05em \textsc{i\kern-.025em b}\kern-.08em
T\kern-.1667em\lower.7ex\hbox{E}\kern-.125emX}}

\def\volumeyear{2016}

\defaultbibliographystyle{SageH}
\defaultbibliography{refs.bib}

\begin{document}

\runninghead{Alexander et al.}

\title{Fragility Measures for Typical Cases}

\author{Robin Alexander*\affilnum{1}, 
        Benjamin R. Baer*\affilnum{2},
        Mario Gaudino\affilnum{3},
        Mary Charlson\affilnum{4},
        Stephen E. Fremes\affilnum{5}, and
        Martin T. Wells\affilnum{6}}

\affiliation{\affilnum{1}School of Medicine, University of St Andrews, Scotland \\
\affilnum{2}School of Mathematics and Statistics, University of St Andrews, Scotland \\
\affilnum{3}Department of Cardiothoracic Surgery, Weill Cornell Medicine, New York, USA \\
\affilnum{4}Department of Medicine, Weill Cornell Medicine, New York, USA \\
\affilnum{5}Division of Cardiac Surgery, Sunnybrook Health Sciences Centre, Ontario, Canada \\
\affilnum{6}Department of Statistics and Data Science, Cornell University, New York, USA \\
}

\corrauth{Robin Alexander, School of Medicine, University of St Andrews North Haugh, St Andrews, KY16 9TF, Scotland}

\email{rnaa1@st-andrews.ac.uk}

\begin{abstract}
    The fragility index is a clinically motivated metric designed to supplement the $p$ value during hypothesis testing. The measure relies on two pillars: selecting individuals or units to have their outcome modified and modifying the outcomes. 
    The measure is interesting, but the unit selection suffers from a drawback which can hamper its interpretation. This work presents the drawback and a method, the stochastic generalized fragility indices, designed to remedy it. Two examples of the causal effect of smoking cessation and electoral outcomes illustrate the method and contribute to the scientific investigation.
\end{abstract}

\keywords{Bush v. Gore, case selection, fragility index, hypothesis testing, interpretability, post-experimental evidence assessment}

\maketitle

\begin{bibunit}

\section{Introduction}
\label{sec:intro}

Null hypothesis significance testing has been the default route for establishing the validity of scientific claims for generations. Throughout this time, the $p$ value has largely been the tool of choice to measure the evidence against a null hypothesis. When properly applied and interpreted, the $p$ value increases the rigor of the conclusions drawn from data \citep{benjamini2021asa}.  Despite their ubiquity, the standard practice involving $p$ values suffers from several shortcomings. Two notable shortcomings are that $p$ values as a post-experimental evidence assessment are widely misunderstood \citep{wasserstein2016asa, benjamini2021asa} and that the default threshold for significance ($p < 0.05$) may not be stringent enough \citep{benjamin2019three, berger1987testing}.

\subsection{The traditional fragility index}
Toward resolving the two shortcomings of $p$ values, \citet{walsh2014statistical} proposed the \emph{fragility index}, a measure that extends a variant proposed by \citet{feinstein1990unit}. \citet{walsh2014statistical} and \citet{feinstein1990unit} focused on analyzing the $2 \times 2$ contingency tables resulting from clinical trials. The (traditional) fragility index is formally defined in Definition~\ref{def:walshfi} \citep{baer2021incidence}.
\begin{definition}
\label{def:walshfi}
    Consider data represented by a $2\times 2$ contingency table where the rows indicate intervention (treatment or control) arms and the columns indicate outcome (event or nonevent) status. The \emph{fragility index} is the minimum number of units whose outcomes must be modified to reverse statistical significance.
\end{definition}
\noindent Note that this definition separates the algorithm and the implicit statistical method in \citet{walsh2014statistical}, as argued by \citet{lin2020factors} and \citet{baer2021incidence}.

The definition relies on a concept of statistical significance. In nearly all applications, this has been taken to correspond to the $p$ value from Fisher's exact test being less than the default cutoff $0.05$. The definition then considers alternative contingency tables wherein each case can have a modified outcome.
Although \citet{walsh2014statistical} defined the fragility index only for initially significant statistical tests, the measure was quickly and modestly extended to initially nonsignificant tests as well, via the so-called \emph{reverse fragility index} \citep{kipp2017vignette, khan2020application}. Although the fragility index was initially motivated by clinical applications, \citet{baer2021incidence} suggest that it should be viewed as a reliable post-experimental evidence assessment that can be applied in all sciences. Toward this end, we use the terminology \emph{unit} in place of \emph{patient} throughout the article.

In many applications, decisions are made on the basis of statistical significance determined by a critical threshold \citep{benjamini2021asa, mayo2021statistics, grimes2024biomedical}.  However, a binary claim of statistical significance is neither
necessary nor sufficient to reach a finding of material significance \citep{poole1987confidence, goodman1999toward, goodman2008dirty, matrixx, greenland2016statistical, lin2023assessing, quatto2025beyond, wang2025role}.  The purpose of the fragility index is to provide a measure of the fragility of a statistical determination.  The fragility index is an interpretable supplement to traditional evidence measures, such as the $p$ value, which is expressed in terms of ``individual units'' instead of probability units. 
The fragility index has been used to re-analyse the principal results in fields across medicine \citep{holek2020fragility}. Researchers in some fields have realized that the statistical significance of practice-changing studies has been based on the outcome of a particular patient (or individual). This is especially troubling when that patient could plausibly have had a different outcome. 

When researchers consider the value of the fragility index to determine whether the conclusion of a study is fragile, their action can be viewed as performing an informal statistical hypothesis test \citep{baer2021samplesize}. To make this clear, we must first update Definition~\ref{def:walshfi}: consider the fragility index to be the \emph{signed} count of outcome modifications. Positive fragility indices correspond to initially significant tests, and negative fragility indices correspond to initially insignificant statistical tests. In a sense, this turns the reverse fragility indices into negative fragility indices. 
This update is summarized in Definition~\ref{def:walshfi2}.
\begin{definition}
\label{def:walshfi2}
    Throughout this article, we consider the fragility index to be a signed variant of Definition~\ref{def:walshfi}, such that positive fragility indices correspond to initially significant tests and negative fragility indices correspond to initially insignificant tests. 
\end{definition}

\begin{figure}
    \centering
    \tikzset{every picture/.style={scale=0.7}}%
%
%
\begin{tikzpicture}[xscale=20]
\draw [thick] (0,0) -- (1,0);

\draw (0,-.2) -- (0, .2);
\node[align=center, below] at (0,-.5){$0$};

\draw (1,-.2) -- (1, .2);
\node[align=center, below] at (1,-.5){$1$};

\draw (.1,-.3) -- (.1, .3);
\node[align=center, below] at (.1,-.5){$0.05$};
\node[align=center, below] at (.1,.91){};

 [9] 0.1011094 0.0635747
\draw (0.1536500,-.2) -- (0.1536500, .2);
\node[align=center, below] at (0.1536500,.81){$-1$};

\draw (0.2237292,-.2) -- (0.2237292, .2);
\node[align=center, below] at (0.2237292,.81){$-2$};

\draw (0.3129707,-.2) -- (0.3129707, .2);
\node[align=center, below] at (0.3129707,.81){$-3$};

\draw (0.4216572,-.2) -- (0.4216572, .2);
\node[align=center, below] at (0.4216572,.81){$-4$};

\draw (0.5484424,-.2) -- (0.5484424, .2);
\node[align=center, below] at (0.5484424,.81){$-5$};

\draw (0.6902888,-.2) -- (0.6902888, .2);
\node[align=center, below] at (0.6902888,.81){$-6$};

\draw (0.8426673,-.2) -- (0.8426673, .2);
\node[align=center, below] at (0.8426673,.81){$-7$};

\node[align=center, below] at (.9,.81){$\dots$};

\draw (0.07111134,-.2) -- (0.07111134, .2);
\node[align=center, below] at (0.07111134,.81){$1$};

\draw  (0.0447549,-.2) --  (0.0447549, .2);
\node[align=center, below] at (0.0447549,.81){$2$};

\draw (0.02718408,-.2) -- (0.02718408, .2);
\node[align=center, below] at (0.02718408,.81){$3$};

\node[align=center, below] at (.01,.81){$\dots$};

\end{tikzpicture}

     \caption{An intuitive depiction of the relationship between fragility indices (on top) and $p$ values (on bottom) when the $p$ value significance threshold is $0.05$. The scale depends on e.g. the sample size and effect size.}
    \label{fig:interval}
\end{figure}
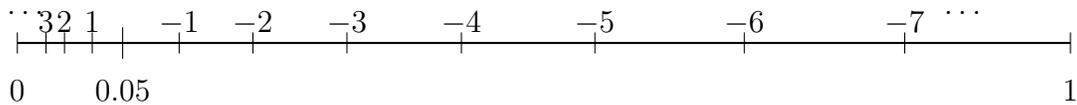

With this improved measure, the fragility index can be neatly treated as a test statistic \citep{lehmann2006testing}. Suppose that we determine statistical significance through the $p$ values. The fragility index being positive is equivalent to the $p$ value being less than the significance threshold. Therefore, the fragility index provides an alternative test statistic for the same rejection region offered by a value $p$ \citep{baer2021samplesize}. This relationship is visualized in Figure~\ref{fig:interval}. However, researchers commonly use the fragility index to go a step further. Indeed, the standard use case of the fragility index is to determine whether the fragility index of a statistically significant test is not ``too low'', for otherwise the trial's statistical conclusion would hinge on a small number of units. This procedure consists of comparing the fragility index with a positive threshold rather than simply $0$ and therefore produces a more stringent statistical test with a lower type I error rate. The need for a more stringent statistical test is further motivated by the lack of reproducibility: reversals of statistical conclusions can be surprisingly common in the medical literature \citep{herrera2019meta}.

The concept underlying the fragility index is largely the same as the $p$ value. Both consider hypothetical outcomes from the same clinical trial. In one case, $p$ values rely on alternative unit outcomes and their distributional impact on test statistics; in the other, the fragility indices directly explore alternative unit outcomes. 

The critiques of the fragility index were reviewed by \citet{baer2021incidence}.
As pointed out in the editorial accompanying the ASA President's Task Force Statement on Statistical Significance and Replicability \citep{kafadar2021statistical}, ``the misuse of a tool should not lead to complete elimination of the tool from the toolbox, but rather to the development of better ways of communicating when it should and should not be used.''

We recommend Section~2 in \citet{baer2021incidence} for an elaborated development of the fragility index. 
Since then, \citet{quatto2025beyond} has proposed a version of the fragility index that has a straightforward mapping to a $p$ value. 
Beyond the formalism of treating the fragility index as a test statistic, the fragility index can be neatly used in a sensitivity analysis \citep{hochstedler2025statistical, liu2025tipping} and operates in a similar manner as the sample breakdown point. 
It can also be understood through a geometric lens using optimal transport theory, which reveals a formal distinction between the fragility index and the p value that goes beyond interpretation.

\subsection{Connections to Robust Statistics and Optimal Transport Theory}

The theoretical basis of statistical science offers several general strategies for dealing with uncertainty in assessing significance \citep{benjamini2021asa}; in this article we will connect some to important concepts from robust statistics and transportation theory.
First we can draw parallels to sample breakdown points, from robust statistics. \citet{baer2021incidence} review an interesting connection between breakdown points and the fragility index quotient, that is, the fragility index divided by the sample size \citep{ahmed2016does}. Then we take a brief look at a geometric abstraction to the fragility index using the Wasserstein distance.

We start by briefly reviewing the breakdown point in non-asymptotic settings.
The breakdown point of an estimator is informally defined as the smallest portion of distributional contamination such that a statistic diverges, that is, breaks down \citep{hampel1968contributions, hampel2011robust, davies2005breakdown}. Distributional contamination can arise in several forms, and here we consider contamination in the form of observation replacement. A principal purpose of the breakdown point is to study the sensitivity of estimators to outliers. Breakdown points have been defined analogously for tests, and we formalize the definition in that context.

Several variants of test breakdown points have been developed \citep{rieder1982qualitative, simpson1989hellinger, jolliffe1993influence}; here we exclusively focus on measures where breaking down means statistical significance reverses. Measures connected to testing breakdown points are compared in Table~\ref{tab:bdp_comparison}.
\citet{ylvisaker1977test} first studied the notion of breakdown for testing and defines the resistance of a test as the smallest fraction of observations that can always determine the test decision regardless of the other observations in the sample. Define $Z$ to be a data sample from $n$ units and $I$ to be a subset of units $\{1, \dots, n\}$. 
The maximum resistance (MR) is formally defined as the minimum cardinality $|I|/n$ such that for all ($\forall$) samples $Z,$ \, there exists ($\exists$) units $I$ for which $\exists$ a modified sample $Z^{\textrm{mod}}$ which reverses significance, where $Z^{\textrm{mod}}$ only differs from the original sample $Z$ for units in $I$ \citep{coakley1994maximum}. This is indicated in the first column of Table~\ref{tab:bdp_comparison}. The maximum resistance tells how robust a test decision is for the least favorable sample and can be viewed as a least upper bound across all samples. 
\citet{coakley1992breakdown} proposed the expected resistance (ER) of a test as a measure of the robustness of its decision through an average across samples with some specified distribution.

\begin{table}
\centering
\begin{tabular}{l||l|l|l|l|l|l}
                         & MR        & ER        & S-BDP/GFI & GFI-SL                 & SGFI       & SGFI-SL                \\
                         \hline 
    Sample $Z$           & $\forall$ & Average   & Given     & Given               & Given     & Given               \\
    Selected units $I$ & $\exists$ & $\exists$ & $\exists$ & $\exists$           & $\exists$ typical & $\exists$ typical            \\
    Modified sample $Z^\textrm{mod}$ & $\exists$ & $\exists$ & $\exists$ & $\exists$ plausible & $\exists$ & $\exists$ plausible
\end{tabular}
\caption{A comparison of methods. The columns represent maximum resistance (MR), expected resistance (ER), sample breakdown point (S-BDP), generalized fragility index (GFI), generalized fragility index with the sufficiently likely construction (GFI-SL), stochastic generalized fragility index (SGFI), and stochastic generalized fragility index with the sufficiently likely construction (SGFI-SL), with cell entries there exists ($\exists$) and for all ($\forall$). }
\label{tab:bdp_comparison}
\end{table}

The maximum and expected resistances are defined across samples rather than for a given sample and are designed to study the abstract properties of statistical tests. On the other hand, in this article, we are interested in studying a given sample $Z$ \citep{donoho1983notion}. \citet{zhang1996sample} introduced the first sample breakdown point (S-BDP) for testing, defined as the minimum $|I|/n$ such that $\,\exists$ units $I$ for which $\exists$ a modified sample $Z^{\textrm{mod}}$ that reverses significance, where $Z^{\textrm{mod}}$ only differs from the original sample $Z$ for units in $I$, for a given sample $Z$. Notice that when the sample $Z$ can be represented as a $2 \times 2$ table, this is precisely the fragility index \citep{walsh2014statistical} in Definition~\ref{def:walshfi} divided by the sample size.

In our view, the fragility index is intended to have a different purpose than breakdown points. Users of the fragility index tend to be interested in the impact of minor perturbations of the data on rejection decisions, for which the sample breakdown point is not suitable. When researchers report a small fragility index, they want the corresponding modifications to plausibly have occurred and not rely on extreme outliers. \citet{baer2021incidence} update the fragility index to explicitly be based on only likely modifications (which correspond to minor perturbations) through the generalized fragility indices (GFIs) with the sufficiently likely construction, as we review in the section \emph{\nameref{sec:methods:stochgen}}.

Each of the measures discussed so far have a commonality. They merely ensure the existence of selected units $I$ that contribute to reversing significance, as seen in the second row of Table~\ref{tab:bdp_comparison}. \citet{donoho1983notion} suggest that this can be a shortcoming and discuss a measure that they call a \emph{stochastic sample breakdown point} which randomly modifies outcomes. The methods we propose, the stochastic fragility index and the stochastic generalized fragility indices, which rely on typical units, will be similar in spirit.

Another interesting perspective of the fragility index can be seen through optimal transport theory. Given two probability distributions the 1-Wasserstein metric quantifies the distance between them based on the minimal cost of transforming one distribution into the other. That is, the 1-Wasserstein distance identifies the smallest amount of probability mass that needs to be redistributed to make one distribution into the other. As the notation is particularly dense, we opt to focus on the intuition here and provide a more verbose exposition in the appendix. In the fragility index setting we can consider $\mu$ to be the empirical distribution of the observed data $Z$ and $N$ to be the set containing the distributions of all data samples for which significance would be reversed. The 1-Wasserstein distance would identify the distribution $Z^\textrm{mod}$ that requires the least changes in individual observations. In the appendix we provide a more detailed summary of how the fragility index can be expressed using optimal transport theory. 

As with the sample breakdown point, the 1-Wasserstein distance does not take into account the plausibility of the nearest distribution, and as such may identify samples where the reversal of significance is reliant on extreme observations.

\subsection{Motivation and roadmap}

As seen in Table~\ref{tab:bdp_comparison}, the fragility index relies on two components that characterize the measure. 
The first is modifying outcomes. \citet{baer2021incidence} thoroughly studied this through the sufficiently likely construction.
The second is selecting the units whose outcomes are to be modified. According to Definitions~\ref{def:walshfi} and \ref{def:walshfi2}, the selected units for the fragility index are the most extreme possible. We will see that these units are atypical, which can hamper their interpretation.


In the section \emph{\nameref{sec:methods}} we introduce methods to generalize and improve the selection of units. In the section \emph{\nameref{sec:examples}} we give real data examples that help develop intuition for the methods. Finally, we summarize and conclude in the section \emph{\nameref{sec:conc}}.

\section{Methods}
\label{sec:methods}

We start with a motivating example. 
Consider the data on the left in Table~\ref{tab:sim_motivating} as arising from a simulated clinical trial. 
The original data has $p$ value $0.49$ and the modified data has $p$ value $0.03$. The data is initially not statistically significant and then becomes significant, with a significance threshold $0.05$. The tables help show that the fragility index is $-7$: there exists seven units for which statistical significance would be reversed had their outcomes been different.

\begin{table}
    \centering
    \begin{tabular}{l||l|l}
        & Event & Nonevent \\
        \hline\hline
        Treatment & 20 & 380 \\
        \hline
        Control & 15 & 385
    \end{tabular}
    \quad
    \begin{tabular}{l||l|l}
        & Event & Nonevent \\
        \hline\hline
        Treatment & 20 & 380 \\
        \hline
        Control & 8 & 392
    \end{tabular} 
    \caption{(Left) Simulated summary statistics; (Right) The modified data which reverses statistical significance}
    \label{tab:sim_motivating}
\end{table}

Because so little information is presented for each unit--only their treatment arm and a dichotomous outcome--we can readily interpret these seven units. Each received the control and had an event. Therefore, we can refine our earlier fragility index interpretation: Seven units who were in the control arm and experienced an event having a different outcome would have reversed significance.

However, interest may not lie in these particular units. There are only $15$ such units among $800$ study participants, resulting in the modified units being atypical enough that they represent only $\frac{15}{800} \times 100\% \approx 1.9\%$ of the study. A user of the fragility index may reasonably be interested in exploring the impact of typical study participants having alternative outcomes. In addition, the units may not be so easily interpretable in studies with more complicated data types.

In this section, we introduce a method that relies on typical units that we call the stochastic fragility indices. 
The method will resolve the interpretation issue motivated above, and we will plainly see through the example in the section \emph{\nameref{sec:examples:pres}} 
that the stochastic fragility indices take into account the rarity of the modified units. This is further illustrated in the section \emph{\nameref{sec:methods:interp}}. 
Note that the motivation underlying the stochastic fragility indices was described as an interesting direction to pursue in \citet{baer2021incidence}.
We then review the generalized fragility indices introduced in \citet{baer2021incidence} and extend the stochastic fragility indices.
in the section \emph{\nameref{sec:examples:adverse}} illustrates that study participants whose outcomes are modified in the generalized fragility index can be acutely atypical when additional covariates are analyzed.
Finally, in the section \emph{\nameref{sec:methods:alg}}.
we introduce an accompanying algorithm in the section \emph{\nameref{sec:methods:alg}}
that is implemented in the open source \texttt{R} package \texttt{FragilityTools} \citep{baer2020fragility}. 

\subsection{The stochastic fragility indices}
\label{sec:methods:stoch}

In this section, we define a method for $2 \times 2$ contingency tables which is based on typical units. We first present a revealing characterization of the fragility index and then leverage it to define the stochastic fragility indices. 

By construction, the fragility index only guarantees the existence of units for which significance would be reversed had their outcomes been different. 
We can see this by Definition~\ref{def:walshfi} since the minimum could be achieved by only one collection of units. Note that for $2 \times 2$ tables, often more than only one collection of units exist. The motivating example based on Table~\ref{tab:sim_motivating} illustrates this: any 7 units could be chosen to reverse significance among the 15 control arm units who experienced an event.

When the fragility index equals $k$, there exists a collection of $k$ units for which significance would reverse had their outcomes been different. However, there are several possible collections of units; when there are $n$ units, there are $\binom{n}{k}$ collections. For example, when $n=800$ and $k=7$, there are more than $4 \times 10^{16}$ unit collections. Of all these collections, the one collection of units guaranteed to have outcomes which can reverse significance can be unusual. 

We now introduce a fragility measure that does not necessarily rely on atypical units. A unit collection has the \emph{reversibility property} if statistical significance would reverse had the units in the collection had different outcomes.
\begin{definition}
    Define the \emph{stochastic fragility index} $\mathit{SFI}_r$ with threshold $r \in [0,1)$ as the minimum $k$ such that more than $r \times 100\%$ of unit collections with cardinality $k$ have the reversibility property. 
\end{definition}
Consider the stochastic fragility index to be signed according to the initial significance of the statistical test, as in Definition~\ref{def:walshfi2}. The stochastic fragility indices can ensure that a substantial portion of possible unit collections can reverse significance, and hence is not forced to rely on atypical units. 

When $r=0$, the stochastic fragility index reduces to the fragility index in Definition~\ref{def:walshfi}. The same holds when $r < 1/\binom{n}{\mathit{FI}}$, where $\mathit{FI}$ is the traditional fragility index.
The stochastic fragility index is undefined when $r=1$. For convenience, we will abuse the notation and write $\mathit{SFI}_1 = \mathit{SFI}_{r}$, when $r=1^{-}$ is the limit as $r$ approaches $1$ from below. In this case, the stochastic fragility index ensures that all cardinality unit collections $\mathit{SFI}_r$ have the reversibility property. We consider this to be a conservative value. Generally speaking, in addition to relying on atypical units, the measure will also rely on atypical units at the opposite extreme.

When $r=1/2$, more than half of the possible unit collections have the reversibility property. That is, most combinations of units have possible outcomes which can reverse significance. We consider this to be a highly interpretable choice and treat it as the default.

The stochastic fragility index generalizes the fragility index to ensure that a particular pattern or collection of units alone do not determine the fragility index result, analogously to the relationship between the stochastic sample breakdown point and the sample breakdown point. 
The stochastic fragility index can equivalently be defined as a quantile with respect to the discrete uniform distribution on the set of unit collections with a given cardinality, as is explored in the section \emph{\nameref{sec:methods:alg}}.
In this sense, the relationship between maximum resistance and expected resistance roughly corresponds to the relationship between fragility indices and stochastic fragility indices.

\subsection{Interpreting the stochastic fragility indices}
\label{sec:methods:interp}

Interpretability is the beating heart of fragility measures. In this section, we study the interpretation of stochastic fragility indices for various choices of $r$. We focus on the case that $\mathit{SFI}_r = 1$ since it is particularly intuitive: unit collections reduce to merely units.

We consider each of the following possible interpretations. The statistical test would not have been significant if: 
\begin{enumerate}
    \item a particular unit had a different outcome,
    \item a typical unit had a different outcome, or
    \item any single unit had a different outcome.
\end{enumerate}

When $r=0$ so that the stochastic fragility index is simply the traditional fragility index, the correct interpretation is the first. By construction, the fragility index only ensures the existence of units that can reverse significance.
When $r=1$, the correct interpretation of the stochastic fragility index is the third. Significance would reverse if any unit had a different outcome, and the word ``single'' considers the multiplicity of units and not which unit.
When $r=1/2$, we consider the correct interpretation of the stochastic fragility index to be the second. By definition, more than half of the units in the study have the property that statistical significance would reverse had their outcome (alone) been different. In our view, more than half of the units in a study cannot be atypical, so one of those units must be typical.

Interpretations for $\mathit{SFI}_r>1$ are analogous except with unit collections instead of individual units. The example in the section \emph{\nameref{sec:examples:pres}}
illustrates an interesting connection between unit collections and the proportion of individual units with a desirable property.

\subsection{The stochastic generalized fragility indices}
\label{sec:methods:stochgen}

The generalized fragility indices directly extend the scope of the fragility index in Definition~\ref{def:walshfi} to arbitrary data types and tests \citep{baer2021incidence}. Let $Z$ be a data frame where the rows represent the units $(1, \dots, n)$ and the columns represent the measurements. For the $2 \times 2$ contingency tables described above, this data frame $Z$ stores the same data, but in a long format. Let the function $m$ be the so-called outcome modifier which inputs a row of $Z$ and outputs the set of \emph{permitted} modifications. 

Writing $\mathcal{R}$ as the rejection region, the generalized fragility indices are formally defined as
\begin{align}
\label{def:genfi}
    \min \,\,\,\, & \| Z - Z^{\mathrm{mod}} \|_{\#} \\
    \text{such that} \,\,\,\, & Z \in \mathcal{R} \,\oplus\, Z^{\textrm{mod}} \in \mathcal{R} \nonumber \\
   & Z^{\textrm{mod}}_{i,} \in \mm (Z_{i,}) \text{ for all } i=1,\dots,n \nonumber
\end{align}
where $\oplus$ is the exclusive-or denoting that $Z$ or $Z^\textrm{mod}$ is in the rejection region (but not both) and $\| \cdot \|_{\#}$ is the norm which counts the number of nonzero rows, i.e. the number of units with modified values for their measurements. 
This definition can be interpreted as a projection of the data, making clear the extreme nature of the generalized fragility index. As in Definition~\ref{def:walshfi2}, consider the generalized fragility indices to be signed according to the initial significance. 
We can readily see that the generalized fragility indices do indeed generalize the fragility index for $2 \times 2$ tables. 

The outcome modifier $m$ needs to be chosen to define a generalized fragility index. We will choose $m$ according to the sufficiently likely construction \citep{baer2021incidence}, which only allows the outcome modifications that have probability at least $q$ for some user supplied $q\in[0,1]$. Thus, the modifier $m=m_q$. We use the notation $\mathit{GFI}_q$ for these fragility measures. The sufficiently likely construction alleviates an issue with the traditional fragility index which strains its interpretation.

In this section, we marry the stochastic fragility indices and the generalized fragility indices to define a fragility measure which both relies on typical cases and permits only plausible modifications.
The method will depend on two parameters: the threshold $r$ that controls how typical the units with reverse significance must be and the sufficiently likely threshold $q$ which controls the plausibility of the modifications. A unit collection has the \emph{reversibility property} if statistical significance would reverse had the units in the collection had different $q$-permitted outcomes, where we recall that an outcome modification is $q$-permitted if it is returned by the modifier $m_q$.
\begin{definition}
\label{def:sgfi}
    Define the \emph{stochastic generalized fragility indices} $\mathit{SGFI}_{r,q}$ with thresholds $r \in [0,1)$ and $q\in[0,1]$ as the minimum $k$ such that
    more than $r \times 100\%$ of unit collections with cardinality $k$ have the $q$-permitted reversibility property.
\end{definition}
Consider the stochastic generalized fragility index to be signed according to the initial significance of the statistical test, as in Definition~\ref{def:walshfi2}.

For data which can be stored in a $2\times 2$ table, notice that a stochastic generalized fragility index with $q=0$ so that any outcome modification is permitted is simply a stochastic fragility index, i.e. $\mathit{SGFI}_{r,0} = \mathit{SFI}_r$. 
The stochastic generalized fragility indices are monotonically nondecreasing in absolute value in both $r$ and $q$ . 

\subsection{An algorithm}
\label{sec:methods:alg}

We now describe an algorithm to approximately calculate a stochastic generalized fragility index and hence also a stochastic fragility index. 
The calculation relies on a different but equivalent presentation of Definition~\ref{def:sgfi}. Let $E^{(q)}_k$ denote whether a uniformly random collection of $k$ units have a permitted outcome modification which reverses statistical significance. Here, the selection of the unit collection is random but each unit measurement is fixed. With this notation, the stochastic generalized fragility index is simply the minimum integer $k$ such that $\mathbb{P}[E^{(q)}_k] > r$, and hence is a quantile.

Thus the value $\mathit{SGFI}_{r,q}$ is the ceiling of the smallest root of $f(k) := \mathbb{P} [E^{(q)}_k] - r$. The function $f$ is nondecreasing since having more units available to receive modifications necessarily increases the probability of reversal. Thus the roots of $f$ are a connected set; for simplicity we will henceforth consider that the root is unique. Note this has not been an issue in practice, except when $r=1$ so that $\mathit{SGFI}_{1,q}$ and any larger count have full probability of reversing significance.

We can readily observe a noisy estimate $\hat{f}$ of $f$ through the following approach. Write $E^{(q)}_k = R^{(q)}(S_k)$ where $S_k$ is a uniformly random sample of $k$ cases and $R^{(q)}$ is a deterministic function which equals True if the cases $S_k$ have permitted outcome modifications which reverse significance and False otherwise. If $R^{(q)}$ was readily available and computable, we may choose $\hat{f}(k) = \frac{1}{B} \sum_{b = 1}^B R^{(q)} (S_{k, b}) - r$ for i.i.d. random case samples $S_{k,b}$ with $b=1, \dots, B$. This is summarised in \ref{alg:reverseprob}.

For other instances, the roots of the noisy estimate $\hat{f}(k) = \hat{\mathbb{E}}[R^{(q)}(S_k)] - r$ can be found, with high probability, using the Polyak-Ruppert averaging algorithm from the stochastic approximation literature \citep{ruppert1988efficient, polyak1992acceleration}. This procedure is displayed in Algorithm \ref{alg:sgfi}. Note that this same algorithm was used to develop a sample size calculation based on fragility index \citep{baer2021samplesize}.

The function $R^{(q)}$ that determines whether significance is reversible can readily be approximated through the greedy algorithm presented in \citet{baer2021incidence}. 
We will run that algorithm with the outcomes fixed for the cases not in the random sample $S_{k, b}$ to determine whether the fragility index is finite or not, i.e. whether significance reversal is possible.

\begin{algorithm}
    \caption{Monte Carlo estimate of the probability of reversing statistical significance}
    \label{alg:reverseprob}
    \begin{algorithmic}[1] 
        \Procedure{ProbabilityReversal}{$k; \alpha, pValue, q, B$}
            \State $\mathit{RevCount} \gets 0$
            \For{\texttt{iter} in 1,\dots,$B$}
                \State $S_k \gets \mathit{Sample}(k)$ \Comment{Randomly sample $k$ cases}
                \State $\mathit{GFI} \gets \mathit{GFIAlgorithm}(q, \alpha, pValue, S_k)$ \Comment{\parbox[t]{.4\linewidth}{Get GFI subject to modifications being permitted only for the cases $S_k$}}
                \State $\mathit{RevCount} \mathrel{+}= I(\mathit{GFI} < \infty)$ \Comment{Increment if $\mathit{GFI}$ is finite}
            \EndFor
            
            \State \textbf{return} $\mathit{RevCount}/B$ \Comment{The proportion of iterations that significance reversed}
        \EndProcedure
    \end{algorithmic}
\end{algorithm}

\begin{algorithm}
    \caption{Algorithm to calculate a stochastic generalized fragility index}
    \label{alg:sgfi}
    \begin{algorithmic}[1] 
        \Procedure{SGFI Calculator}{$q,r,\alpha,\text{function } pValue, B$}
            \State $\hat{f}(k) \gets \text{ProbabilityReversal} (k; \alpha, pValue, q, B)-r$
            \State $\mathit{SGFI} \gets \text{FindRoot}(\hat{f})$ \Comment{Get the root of $\mathbb{E}[\hat{f}]$ using Polyak-Ruppert averaging}
            \State \textbf{return} $\mathit{SGFI}$
        \EndProcedure
    \end{algorithmic}
\end{algorithm}

In summary, we approximately calculate a stochastic generalized fragility index through a stochastic root finding algorithm, Monte Carlo estimates, and a greedy approximation.

\section{Examples}
\label{sec:examples}

In this section we review two interesting examples of the fragility measures defined earlier. 
The first example illustrates a typical data analysis that makes use of each fragility measure presented. Part of the example illustrates that the units chosen by a generalized fragility index to have their outcome modified are particularly atypical in the presence of a continuous covariate.
The second example offers intuition for the stochastic fragility indices by connecting the proportion of units with a desirable property to the unit collections in the stochastic generalized fragility index.
Each example can be reproduced using scripts in the \texttt{R} package \texttt{FragilityTools} \citep{baer2020fragility}. 

\subsection{Modelling an adverse event}
\label{sec:examples:adverse}

The NHEFS, an observational study, and corresponding data set were relied on and analyzed throughout the causal inference textbook by \citet{hernan2020causal}. 
(The acronym stands for \emph{National Health and Nutrition Examination Survey Data I Epidemiologic Follow-up Study}.) 
The data set is a sample of 1629 cigarette smokers aged 25-74 years who had a baseline in the 1970s and then a follow up a decade later. The purpose of the study was to investigate the relationships between clinical, nutritional, and behavioral factors and several adverse events. In this section, we will study the relationship of smoking cessation between baseline and 1982 (the exposure) and death by 1992 (the endpoint).

\begin{table}
    \centering
    \begin{tabular}{l||l|l}
        & Death & Survival \\
        \hline\hline
        Quit smoking & 102 & 326 \\
        \hline
        Continued smoking & 216 & 985
    \end{tabular}
    \caption{Summary statistics from NHEFS}
    \label{tab:nhefs}
\end{table}

\subsubsection{Fragility indices for \texorpdfstring{$2 \times 2$}{2 x 2} tables}

An early analysis of the relationship could leverage the data in Table~\ref{tab:nhefs}.
The odds ratio for smoking cessation is $1.43$, indicating that quitting smoking may increase the risk of death.
Fisher's exact test for whether smoking cessation is associated with death has $p$ value $0.01$. Taking the significance cutoff as the default $0.05$, we would conclude that that smoking cessation is associated with an increased risk of death. The fragility index is $6$, revealing that a few particular units would need to have modified outcomes to reverse significance. The incidence fragility indices reveal that the outcome modifications were rather likely: any $q\in[0, 0.76)$ gives the same value \citep{baer2021incidence}. 

The $6$ units whose outcomes were modified each from quit smoking and died to quit smoking and survived. These units are atypical in the study and comprise only $6\%$ of the study participants. The stochastic fragility indices ensure that units across the study can contribute to significance reversing. Here we find $\mathit{SFI}_{0.5} = 22$, showing that a typical collection of 22 units having outcome modifications would reverse statistical significance.
Notice that the representation in the previous section does not hold (i.e. $22 \not\approx \frac{6}{0.06}$) since more than just cases where people quit smoking and died can contribute to the reversal of significance.

\subsubsection{Generalized fragility indices}
The results of the previous section could naively be interpreted as suggesting that smoking cessation is harmful. However, determining a causal relationship requires controlling for confounders. 
Possible confounders include years smoked, sex, race, weight, etc. 
It is important to control for confounders because the association described previously may be spurious if the distribution of years smoked differs between study arms. 
For the purpose of better illustrating the stochastic fragility indices, we will treat years smoked as the only confounder. The arguments in this section will still hold in the presence of more confounders, but the visualizations will be more complicated. 
The $p$ value for whether smoking cessation is associated with death controlling for years smoked is $p=0.41$ in a logistic regression, with adjusted odds ratio $1.13$. This is initially insignificant with the usual threshold $0.05$.

The traditional fragility index and the incidence fragility indices do not allow confounders so we now consider generalized fragility indices.
The generalized fragility index permitting any modification (i.e. having $q=0$) is $-10$ so that at least ten units must have their death status modified to reverse significance. 

\begin{figure}
    \centering
    \includegraphics[scale=.7]{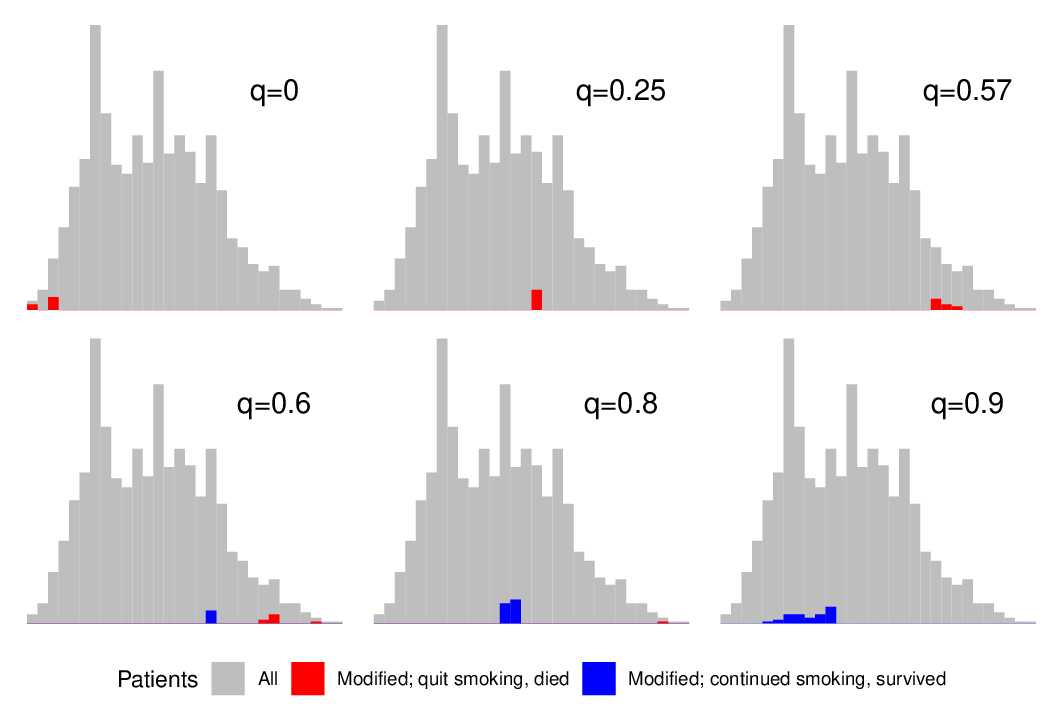}
    \caption{Histograms for the confounder, number of years smoked. The grey bins correspond to all cases in the study (min=1, max=64), and the colored bins correspond to cases selected to have their outcome modified by the generalized fragility index.}
    \label{fig:hnefs_gfi}
\end{figure}

This generalized fragility index belies the smoking years of the ten selected units. 
Clinicians may imagine that these units are typical and so are not consistently in especially poor or excellent condition.
However, the generalized fragility index seeks to reverse significance with the fewest outcome modifications and hence relies on atypical units for which their outcome modification can have an extreme impact. The top-left pane in Figure~\ref{fig:hnefs_gfi} shows the distribution of years smoked for the selected units relative to all of the units in the study. The selected units (in red) each quit smoking and died, as for the traditional fragility index in the previous subsection. The selected units also each have low years smoked since modifying these units' outcomes most alters the $p$ value. 

The remaining panes show the same for different values of $q$. As $q$ grows larger than $0$, the selected units having increasingly higher years smoked since the units with lower years smoked no longer have an outcome modification permitted. When $q=0.57$ the years smoked of the selected units reaches the right tail in the distribution of all the units. After, at $q=0.6$, units who did not quit smoking and survived (in blue) begin to have their outcome modified. The years smoked of these selected units is large. As $q$ grows further, the selected units have lower years smoked. When $q=0.9$, the years smoked of the selected units reaches the left tail of the distribution of all the units. 

We see that the most atypical units with permitted modifications will be selected to modify their outcome. 

\subsubsection{Stochastic generalized fragility indices}

\begin{figure}
    \centering
    \includegraphics[scale=.8]{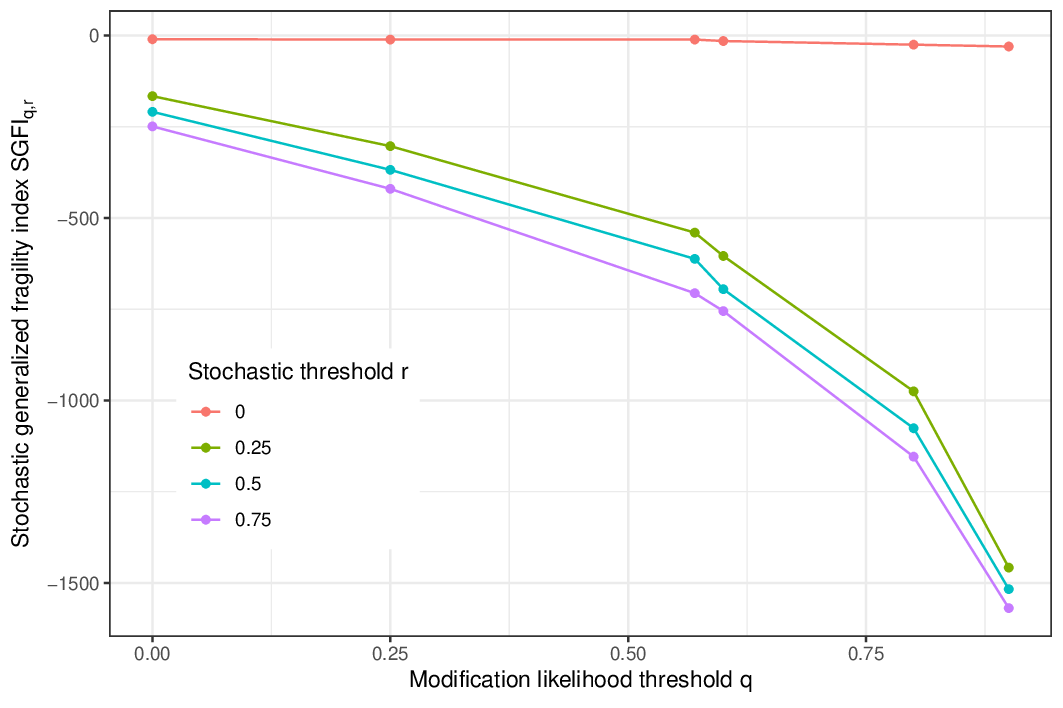}
    \caption{The stochastic generalized fragility indices for various choices of stochastic threshold $r$ and sufficiently likely threshold $q$.}
    \label{fig:nhefs_sgfi}
\end{figure}

The stochastic fragility indices resolve this shortcoming by ensuring that outcome modifications of typical units can reverse significance. Figure~\ref{fig:nhefs_sgfi} visualizes the stochastic generalized fragility indices corresponding to the test that controls the years of smoking confounder. Notice that the stochastic generalized fragility index is monotonically decreasing in both $r$ and $q$.

When $q=0.9$ so that only very likely outcome modifications are allowed, the generalized fragility index is $-30$ (indicated in red at the top of the figure). When the stochastic threshold $r$ grows beyond $0$ to $0.25, 0.5,$ or $0.75$, the stochastic generalized fragility indices are much larger: they are $-1458$, $-1517$, and $-1569$, respectively.
Therefore, for example, $1517$ units must be selected to ensure that typical (i.e. more than half of) unit collections can reverse significance with only very likely modifications (i.e. modifications which have likelihood at least $0.9$). Whether this number is low enough to suggest that the statistical significance decision is fragile depends on the preferences of the researcher. 

In general, the choices $r=0.25, 0.5,$ and $0.75$ produce similar stochastic generalized fragility indices for each value of the sufficiently likely threshold $q$. 
Notice that these values are large because few units have allowed outcome modifications, which can contribute to reversing significance when $q$ is large.

\subsection{Presidential election}
\label{sec:examples:pres}

We believe that fragility measures are applicable beyond clinical trials and statistical hypothesis testing. For example, a generalized fragility index can be used to formalize a critique of the electoral college in United States presidential elections \citep{dixon1950electoral}. In the 2000 presidential race of Bush versus Gore, Bush won the election due to the Electoral College, with a final tally of 50,999,897 votes for Gore and 50,456,002 votes for Bush. Additionally, 92,875,537 eligible voters did not vote for either; for convenience, we call these nonvoters, despite some voting for a third party. Note that in practice the decision of who will be the US President can depend on more than just votes due to the possibility of ballot recounting, judicial review, faithless electors, etc. 

This data type is reminiscent of that for meta-analyses, where states are instead studies. This is because individual votes within each state are tallied and then the electoral votes are, in almost all states, all cast towards the candidate with the most votes in that state.

In this section, we study this example and make an interesting connection to the generalized fragility indices and stochastic generalized fragility indices. A particularly interpretable representation for the stochastic fragility index (with $r=0.5$) is found.

\subsubsection{\emph{Generalized fragility index}}

More detailed data by state reveal that Gore would have won the election had 538 non-voters in Florida instead voted for Gore \citep{florida2000}. This number was widely broadcast at the time \citep{purdum2000counting} and was used to argue that the election decision was fragile.

Even though this example does not involve a statistical test, it demonstrates fragility of a decision through outcome modifications, and hence is a kind of fragility measure. To formalize the connection to a generalized fragility index, we now make the elements previously outlined in the section \emph{\nameref{sec:methods:stochgen}}
more concrete. 
Let $Z$ be a data frame with 2 columns denoting the State and the vote (either `Bush', `Gore', or `Neither') and with a row for each eligible American voter. 
Let the outcome modifier $m$ be unrestricted among nonvoters but fully restricted among voters so that nonvoters can have their vote modified to `Bush', `Gore', or `Neither' but the vote for those who already committed to Bush or Gore cannot be modified. Note that this modifier $m$ is not chosen according to the sufficiently likely construction. 
Finally, let the decision $\mathcal{R}$ indicate whether Bush won the election or Gore won the election. 
This generalized fragility index is thus the smallest count of vote modifications to nonvoters necessary to reverse the outcome of the US election. 

We now show that the circumstances of the election outcome lead to a helpful simplification. 
According to the vote margins in each state, Florida, New Hampshire, and Nevada are the only red states that could flip to blue states with a moderate amount of additional Gore votes. Any one of these three states going blue would have made Gore win the election; however, New Hampshire and Nevada were both much smaller than Florida and required many more Gore votes to turn blue. Therefore, we may reasonably make the simplifying assumption that a moderate sized collection of eligible voters in the US can only reverse the US Presidential race by reversing the result of the Florida race. 
Thus, the generalized fragility index is the smallest count of vote modifications to \emph{Florida} nonvoters necessary to reverse the outcome of the US election, which is the $538$ number cited earlier. 

\subsubsection{Stochastic generalized fragility index}

In this section, we turn our attention to the stochastic generalized fragility indices. We focus on the case $r=0.5$ as it is reasonable and will produce a particularly intuitive representation. 

In Table~\ref{tab:nhefs} we provide a summary of the voting grouped by Florida versus the rest of the United States \citep{eac2000voter}.

\begin{table}
    \centering
    \begin{tabular}{l||l|l|l}
        & Voted & Not Voted & Total Registered Voters \\
        \hline\hline
        Florida & 5,963,110 & 2,789,607 & 8752717\\
        \hline
        Not Florida & 99,623,164 & 48,045,430 & 147,668,594\\
        \hline \hline
        \multicolumn{3}{l}{} & \multicolumn{1}{|l}{156,421,311} \\
    \end{tabular}
    \caption{Summary of voting results from the 2000 US presidential election.}
    \label{tab:nhefs}
\end{table}


Since $538$ Floridian nonvoters would need to vote for Gore to reverse the result of the Florida race, we seek the lowest count $\mathrm{SGFI}_{0.5}$ such that a random collection of
$\mathrm{SGFI}_{0.5}$ registered American voters is more likely than not to include
$538$ Florida registered voters who did not cast a vote for President in 2000.
Accordingly, we seek the lowest value $\mathrm{SGFI}_{0.5}$ such that
\begin{equation}
  \frac{1}{2}
  < 
  \mathbb{P}\left[
    538 \leq \mathrm{HyperGeometric}\!\left(156421311, 2789607, \mathrm{SGFI}_{0.5}\right)
  \right],
\end{equation}
where $156{,}421{,}311$ is the number of registered U.S.\ voters and $2{,}789{,}607$
is the number of Florida registered voters who did not cast a presidential vote.


Thus $\textit{SGFI}_{0.5}$ approximately satisfies that $\mathrm{median} \left( \mathit{HG} \right) = 538$, where $\mathit{HG} = \mathrm{HyperGeometric} \left( 156421311, 2789607, \textit{SGFI}_{0.5} \right)$.
Due to the large values of the first two parameters, this hypergeometric distribution is approximately equal to a Binomial distribution with parameters $\textit{SGFI}_{0.5}$ and $\hat{p}_{\text{FL}}$, where $\hat{p}_{\text{FL}}=2789607/156421311 \approx 0.018$ is the empirical probability of selecting a Floridian nonvoter among all American eligible voters \citep{blitzstein2019introduction}. Thus the Binomial distribution mean $\textit{SGFI}_{0.5} \hat{p}_{\text{FL}}$ is approximately equal to the Hypergeometric median, and we have that 
\begin{equation}
    \textit{SGFI}_{0.5} \approx  \frac{\mathit{GFI}}{\hat{p}_{\text{FL}}} \approx \frac{538}{0.018} \approx 29889.
\end{equation}

Therefore, $29{,}889$ eligible American voters need to be selected to ensure that a typical collection includes enough nonvoters to overturn the US election. 
These tens of thousands of random \emph{American eligible voters} revealed by the stochastic generalized fragility index are in sharp contrast to the $538$ \emph{Floridian nonvoters} revealed by the generalized fragility index. The former is representative of all Americans, but the latter exclusively concerns Floridians, despite the generalized fragility index nominally involving all Americans.

Whether this number is low enough to suggest that the election decision is fragile depends on the preferences of the researcher. The authors feel that the number is surprisingly low given the roughly $156$ million eligible American voters.

This stochastic generalized fragility index is simply an up-weighted generalized fragility index, directly taking into account the rarity $\hat{p}_\text{FL}$ of eligible voters who must be selected to reverse the result of the election. 
The representation is reminiscent of the RIR (Robustness of an Inference to Replacement) method, but the denominator probabilities are distinct \citep{frank2021hypothetical, baer2022three, frank2022response}.
Recall that this representation hinged on Florida being the only state of interest for reversing the election due to the electoral college; an analogous property will not generally hold for statistical tests, as we explore in the next section.

\section{Conclusion}
\label{sec:conc}

We believe there is a promising future for statistics based on unit counts. They are broadly interpretable to medical researchers and others from varied backgrounds. The fragility index is an interesting addition to a statistician's toolkit, alongside the other fragility measures developed here.

The measures developed here represent a substantial departure from earlier formulations of the fragility index. 
\citet{walsh2014statistical} and \citet{feinstein1990unit} defined the fragility index purely through existence: 
the minimum count of units whose outcomes could be modified to reverse significance, 
without regard for which units are selected or how plausible the modifications are. 
This limitation can be understood through two complementary lenses. 
The first is combinatorial: as summarized in Table \ref{tab:bdp_comparison}, 
the progression from existential to typical unit selection parallels the classical progression from sample breakdown points to stochastic breakdown points \citep{donoho1983notion}, 
applied here to post-experimental evidence assessment. 
The second draws on optimal transport theory: the fragility index quotient equals the $1$-Wasserstein distance between the observed empirical distribution and the nearest nonsignificant distribution, giving fragility a geometric interpretation as minimal mass transport in the space of empirical distributions. 
In both views, the traditional fragility index identifies the most extreme perturbation without accounting for plausibility or typicality. 
The stochastic generalized fragility indices address this by constraining both frameworks, 
requiring that modifications occur across typical units with sufficiently likely outcome changes, rather than through the most extreme available perturbation.  

The stochastic generalized fragility indices complete the foundational methodological development of the fragility index. While the sufficiently likely construction permits only plausible outcome modifications to ensure that modifications are realistic, the stochastic property forces modifications to occur across typical units rather than only in particular units. The stochastic generalized fragility indices study and improve both the unit selection and the outcome modification selection which defines fragility measures. 

Through the examples looking at adverse events and the 2000 US presidential election, we illustrate the deficiencies of the traditional and generalized fragility indices and the resolution to the deficiencies through the stochastic fragility indices. 
First, we saw that the fragility index tends to focus on (or ``pick on'') a single strata of cases when selecting cases for outcome modification, such as cases in a particular arm of a study (in the section \emph{\nameref{sec:methods}}) 
or voters in the particular state Florida (in the section \emph{\nameref{sec:examples:pres}}). 
Second, we saw that the selected cases being ``picked on'' tend to be in the most atypical strata, such as voters in Florida having the closest race (in the section \emph{\nameref{sec:examples:pres}}) 
or explanatory variables being in the tail of their distribution (in the section \emph{\nameref{sec:examples:adverse}}). 
Both deficiencies are addressed by the stochastic generalized fragility indices by ensuring that all cases may contribute to reversing significance. 
Throughout, we saw that the stochastic generalized fragility indices take into account the rarity of cases whose outcomes must be modified to reverse significance in the generalized fragility index, although this is particularly salient in the example in the section \emph{\nameref{sec:examples:pres}}.

The interpretation of the stochastic generalized fragility indices were illustrated in the previous section focused on interpreting the stochastic fragility index in the section \emph{\nameref{sec:methods:interp}} 
and throughout the examples in the section \emph{\nameref{sec:examples}}.
The former gave a simple and clear interpretation of stochastic fragility indices, which are defined using data in a $2\times 2$ table. The choice of the stochastic index $r$ determines a particular case, or any single case can reverse significance when $\mathit{SFI}_r=1$, or any intermediate representativeness of cases. When $r=0.5$ for general data types, the stochastic fragility index can simply be interpreted as reporting the number of typical case outcome modifications needed to reverse significance.

All fragility measures are based on choosing permitted outcome modifications that have the largest impact on significance. Put differently, they all rely on an adversarial choice of the outcome modification \citep{lowd2005adversarial}. In view of Table~\ref{tab:bdp_comparison}, this is due to the ``$\exists$'' in each entry of the last row.
Future work that introduces a new category of measures that deviates from this could be interesting and may bridge the gap between fragility measures and $p$ values. Randomly choosing outcome modifications may work for fragility indices with initially significant tests but not in general.

\section*{Acknowledgements}

The authors gratefully thank Apurva Dixit and Derrick Tam for developing an early version of the stochastic fragility indices, alongside SEF. 

\section*{Funding}

BRB's research was partially supported by NIH grant R61/R33 NS120240. MTW’s research was partially supported by NIH grants R01 GM135926 and 1P01AI159402. MTW and MC’s research was partially supported by Patient-Centered Outcomes Research Institute grant IHS-2017C3-8923. The funding sources had no direct role in this paper.

\putbib
\end{bibunit}

\begin{bibunit}
\appendix 
\section*{Appendices}

\section{The Fragility Index as an Optimal Transport Problem}
\label{sec:app:transport}

In the section \emph{\nameref{sec:intro}}, we described the connection between the fragility index and optimal transport theory at an intuitive level. Here we provide the formal details, showing that the fragility index quotient \citep{ahmed2016does} can be expressed as a 1-Wasserstein distance.

\subsection{The 1-Wasserstein distance}

The 1-Wasserstein distance, also known as the \emph{Earth Mover's Distance}, quantifies the distance between two probability distributions by the minimal cost of transforming one into the other \citep{villani2008optimal, peyre2019computational}.

Let $\mu$ and $\nu$ be two probability measures defined on a metric space $(\mathcal{X}, d)$. The 1-Wasserstein distance between them is
\[
W_1(\mu, \nu) = \inf_{\gamma \in \Gamma(\mu, \nu)} \int_{\mathcal{X} \times \mathcal{X}} d(x, y) \, d\gamma(x, y),
\]
where $\Gamma(\mu, \nu)$ denotes the set of all couplings (joint distributions) with marginals $\mu$ and $\nu$, and $d(x, y)$ is the cost of transporting mass from $x$ to $y$.

When $\mu = \frac{1}{n} \sum_{i=1}^n \delta_{x_i}$ and $\nu = \frac{1}{n} \sum_{i=1}^n \delta_{y_i}$ are empirical measures on the real line with $d(x, y) = |x - y|$, this simplifies to
\[
W_1(\mu, \nu) = \frac{1}{n} \sum_{i=1}^n |x_{(i)} - y_{(i)}|,
\]
where $x_{(i)}$ and $y_{(i)}$ denote the sorted values of the respective samples.

\subsection{The Wasserstein fragility index}

Let $\mathbb{P}_0$ denote the empirical distribution of the observed data $Z$, supported on $\{0,1\}^n$ for binary outcomes. Define the set of distributions that yield nonsignificant results as
\[
\mathcal{F}_\alpha = \left\{ \mathbb{P} : p(T(\mathbb{P})) > \alpha \right\}.
\]
We define the \emph{Wasserstein Fragility Index} (WFI) as
\[
\text{WFI}(\mathbb{P}_n) := \inf_{\mathbb{P} \in \mathcal{F}_\alpha} W_1(\mathbb{P}_n, \mathbb{P}),
\]
which represents the minimal mass transport needed to reach a nonsignificant distribution from the observed one. We can modify this in the usual way to accommodate either significant or insignificant data sets. 

In the discrete case with binary outcomes $X_i \in \{0,1\}$ and equal unit cost for flipping an outcome (i.e., $d(x,y) = |x - y|$), we have
\[
W_1(\mathbb{P}_n, \mathbb{P}') = \frac{k}{n},
\]
where $k$ is the number of label flips and $n$ is the sample size. It follows that
\[
\text{WFI}(\mathbb{P}_n) = \frac{\text{FI}(Z)}{n},
\]
where $\text{FI}(Z)$ is the fragility index in Definition~\ref{def:walshfi}. That is, the fragility index quotient equals the 1-Wasserstein distance to the nearest nonsignificant distribution. This gives a geometric interpretation of fragility: the transition from statistical significance to nonsignificance is captured by the minimal transport cost in the space of empirical distributions.

\putbib
\end{bibunit}

\end{document}